\documentclass[conference]{IEEEtran}
\usepackage{graphicx}
\usepackage{lipsum}
\usepackage{amsmath}
\usepackage{multirow}
\usepackage{booktabs}
\usepackage[bookmarks=false]{hyperref}
\usepackage{caption}
\usepackage{subcaption}
\usepackage{xcolor}
\usepackage{endnotes}
\let\footnote=\endnote
\begin{document}

%%%%%%%%%%%%%%%%%%%%%%%%%%%%%%%%%%%%%%%%%%%%%%%%%%%%%%%%%%%%%%%%%%%%%%%%%%%%%%%%
\title{Convolutional-Recurrent Neural Networks on Low-Power Wearable Platforms for Cardiac Arrhythmia Detection}

%%%%%%%%%%%%%%%%%%%%%%%%%%%%%%%%%%%%%%%%%%%%%%%%%%%%%%%%%%%%%%%%%%%%%%%%%%%%%%%%
\author{
\IEEEauthorblockN{Antonino Faraone}
\IEEEauthorblockA{ETH, Z\"{u}rich, Switzerland\\
afaraone@student.ethz.ch}
\and
\IEEEauthorblockN{Ricard Delgado-Gonzalo}
\IEEEauthorblockA{CSEM, Neuch\^{a}tel, Switzerland\\
ricard.delgado@csem.ch}
}

%%%%%%%%%%%%%%%%%%%%%%%%%%%%%%%%%%%%%%%%%%%%%%%%%%%%%%%%%%%%%%%%%%%%%%%%%%%%%%%%
\maketitle

%%%%%%%%%%%%%%%%%%%%%%%%%%%%%%%%%%%%%%%%%%%%%%%%%%%%%%%%%%%%%%%%%%%%%%%%%%%%%%%%
\begin{abstract}
Low-power sensing technologies, such as wearables, have emerged in the healthcare domain since they enable continuous and non-invasive monitoring of physiological signals. In order to endow such devices with clinical value, classical signal processing has encountered numerous challenges. However, data-driven methods, such as machine learning, offer attractive accuracies at the expense of being resource and memory demanding. In this paper, we focus on the inference of neural networks running in microcontrollers and low-power processors which wearable sensors and devices are generally equipped with. In particular, we adapted an existing convolutional-recurrent neural network, designed to detect and classify cardiac arrhythmias from a single-lead electrocardiogram, to the low-power embedded System-on-Chip nRF52 from Nordic Semiconductor with an ARM's Cortex-M4 processing core. We show our implementation in fixed-point precision, using the CMSIS-NN libraries, yields a drop of $F_1$ score from 0.8 to 0.784, from the original implementation, with a memory footprint of 195.6~KB, and a throughput of 33.98~MOps/s.
\end{abstract}

%%%%%%%%%%%%%%%%%%%%%%%%%%%%%%%%%%%%%%%%%%%%%%%%%%%%%%%%%%%%%%%%%%%%%%%%%%%%%%%%
\section{Introduction}
The recent developments in the field of Deep Learning (DL) gave an important boost to the field of healthcare and biomedical engineering~\cite{Miotto2017}. The unprecedented accuracy enabled by Deep Neural Networks is progressively overwhelming algorithms based on classical signal processing for many application scenarios due to the availability of large datasets and raw computational power. On another side, wearable devices are showing a high potential in the healthcare domain~\cite{Dunn2018} as they enable continuous and non-invasive monitoring of vital parameters, prompt detection of disorders and diseases, and an early detection of emergencies. Wearable devices have achieved great success both for personal healthcare management (\textit{e.g.}, wristbands~\cite{Nelson2016}, smart-vests~\cite{Schneegass2017}) and for support to clinical treatments~\cite{Mombers2016}. The benefit coming from the implementation of DL techniques in such wearables would be a game-changer for the whole sector~\cite{Rajpurkar2017}, but it is systematically hindered by the hardware limitations of such devices, namely limited computational power, memory, and battery life~\cite{Chen2017}.

For these reasons, wearable applications that want to exploit neural networks generally offload such computations to a remote cloud server that collects data produced by the resource-limited sensors, process them on high performance hardware, and return back the results to the user or, potentially, to a medical doctor and emergency services~\cite{Rashid2017}. This workaround requires a reliable connection to the cloud server, introduces latency issues on real-time applications, and raises privacy concerns~\cite{Horvitz2015}. Moreover, transmitting signals at a high sampling rate from sensors to edge devices is extremely demanding in terms of energy. This becomes unsustainable in the case of battery-powered devices for continuous monitoring of electrocardiograms (ECG) or electroencephalograms (EEG). To overcome the above obstacles, a suitable and emerging solution is to bring data processing as close as possible to the devices that produced it. This means, for instance, to perform expensive computations on edge devices like single-board computers~\cite{Lee2013,Li2018}, mobile GPUs, dedicated hardware~\cite{Linares-Barranco2019,Gao2019}, and smartphones~\cite{Ignatov2018}. 

In this paper, we focus on the inference of neural networks running on microcontrollers and low-power processors, which wearable sensors and devices are generally equipped with. We chose as use case the detection and classification of cardiac arrhythmias. Arrhythmias are cardiac irregularities of heart beats that can lead to severe health complications~\cite{Bennett2012}. There are several categories of Arrhythmias whose detection and diagnosis is generally performed by specialists in cardiology via analysis of ECGs. We extend the work of Van Zaen \textit{et al.}~\cite{VanZaen2019}, a convolutional-recurrent neural network architecture for atrial fibrillation detection, trained on the dataset provided for the 2017 Computation in Cardiology Challenge~\cite{Clifford2017}. This network achieves an F1 score of 0.81 for detection of atrial fibrillation, and has been validated on ECG acquired with sensors from a smart vest~\cite{Chetelat2015}. This network serves as a baseline for our work. Our work focuses on the trade-offs between model complexity and performance drops. Major attention is paid on architectural changes to reduce memory footprint and operations count of the model.

The paper is structured as follows. In Section~\ref{sec:mat}, we introduce the software libraries, hardware, and data employed to train and evaluate our embedded neural network. In Section~\ref{sec:arch}, we present the optimized NN architecture as well as the steps performed to optimize it for the deployment on the target SoC. Then, in Section~\ref{sec:results}, we analyze our proposed NN in terms of memory footprint, execution time, and overall operation count and throughput when running into the target SoC. Finally, we conclude by outlining the benefits and limitations of our approach and setting the direction for further work.

%%%%%%%%%%%%%%%%%%%%%%%%%%%%%%%%%%%%%%%%%%%%%%%%%%%%%%%%%%%%%%%%%%%%%%%%%%%%%%%%
\section{Materials and Methods}\label{sec:mat}
In this section, we first present and discuss the main software tools that we leverage in our implementation. Then, we present the target hardware platform and provide an overview of its technical specifications and limitations. Finally, we introduce the dataset that was used during training.

\subsection{Software Tools}
CMSIS\footnote{\url{https://github.com/ARM-software/CMSIS_5}} is a software library that provides a hardware abstraction layer for ARM Cortex-based processors. It includes a DSP library and, from version v5, a set of routines to deploy neural networks on Cortex microcontrollers named CMSIS-NN~\cite{Lai2018}. It supports a basic range of layer typologies, namely convolutional layers, dense layers, and pooling layers, various activation functions, including $\tanh$ and sigmoid, and a modified version of Softmax that works with power of 2 instead of $\textrm{e}$. In order to reduce memory footprint and speedup computations, CMSIS-NN employs fixed-point quantization, consisting in representing weights and activations as 8 or 16~bit signed integers in $Qn.m$ format, where $n$ and $m$ are respectively the number of bits allocated for the integer and fractional part. the Q-format for each weight and activation must be chosen a-priori by analyzing their range of values. If $B$ is the number of bits allocated for a variable $v$, excluding the sign bit, to convert it to the corresponding $Qn.m$ representation, the following steps are performed:
\begin{equation}\label{eq:quant}
\begin{aligned}
v&=\mathrm{round}(v\cdot 2^m) \\
v&=\mathrm{clip}(v,[-2^B,2^B-1]) \\
\end{aligned}
\end{equation}
The advantage of such representation in terms of computation complexity is that the computations do not require the Floating Point Unit (FPU), as all numbers are actually treated as as integers. Furthermore, Cortex-M4 and M7 processors support SIMD Instructions (Single Instruction Multiple Data) capable of operating simultaneously on multiple 16~bit integer operands.

For each layer, two more parameters have to be fixed, namely the shifts for the bias and the output. If weights are in $Qx.y$ format, and inputs are in $Qa.b$ format, the product between those two tensors will have $Q(x+a).(b+y)$ format. Therefore, if the format of the biases does not match it, it is necessary to apply a shift to them, that must be calculated a-priori during the network implementation. Finally, the output must be shifted to match the format of the input of the following layer. CMSIS-NN provides fast versions of convolutional layers that employ further optimization tricks but that impose the constraint of having input channels multiple of 4 and output channels multiple of 2. 

\subsection{Hardware Platform}
Our target hardware platform is the nRF52832 SoC from Nordic Semiconductor. It is powered by an ARM Cortex-M4 MPU clocked at 64~MHz, equipped with 64~KB of RAM and 512~KB of FLASH memory. It targets low-power Bluetooth applications like Internet-of-Things (IoT) and medical wearable devices. The advertised supply current is 3.7~mA (running from FLASH, using internal DC/DC, 3~V supply voltage), while this figure drop to just 0.3~$\mu$A in OFF mode without RAM retention. The platform includes an FPU and supports SIMD instructions, which are heavily used in CMSIS-NN to speedup matrix multiplications and convolutions.

\subsection{Dataset}
The chosen dataset consists of 8,528 samples of single-lead ECG signals, used as reference dataset for the Computing in Cardiology 2017 Challenge. The raw signals are sampled at 300~Hz and have a variable duration between 9 and 60 seconds~\cite{Clifford2017}. Each sample is labeled over four classes: Normal Rhythm, Atrial Fibrillation, Noise, and Other Rhythm. Classes are unbalanced (with strong predominance of normal rhythms) and weakly labeled, meaning that each label is associated to the whole recording, thus we have no information about the exact samples range where arrhythmia occurs. Since the official test set used for the competition has not been released, we used instead a subset of 1,528 signals extracted from the dataset, striving to keep the same proportion between classes. The remaining 7,000 samples have been used for training.

Several pre-processing steps are applied to the dataset as described in~\cite{VanZaen2019}. First, it is filtered using a Butterworth band-pass filter with passband between 0.5 and 40~Hz. Then, we resample each signal at 107~Hz in order to reduce the workload in the final implementation and match the sampling frequency of the acquisition device used for internal demonstrations. The records are normalized and, before feeding them to the network, they are split into windows of 256 samples with 50\% overlap. If signals have a number of samples not divisible by 256, a number of samples are discarded from the beginning and the end of the sequence by applying a random offset to the first window.

%%%%%%%%%%%%%%%%%%%%%%%%%%%%%%%%%%%%%%%%%%%%%%%%%%%%%%%%%%%%%%%%%%%%%%%%%%%%%%%%
\section{Neural Network}\label{sec:arch}
In this section, we describe the modifications that we implemented in the NN from~\cite{VanZaen2019} in order to be able to run into our target platform.

\subsection{Architecture}
The NN can be decomposed in two parts. Each window is first processed through a sequence of 7 convolutional layers of size 5, each followed by an average pooling layer with size and stride equal to 2. The number of channels is kept multiple of 8 in order to exploit the speedup from the optimized convolutional kernels of CMSIS-NN. A global averaging pooling layer is applied after the last layer. The output of the convolutional part is a set of 128-dimensional tensors, one for each window, which is then fed into a Gate Recurrent Unit (GRU) with 64 hidden units. Moreover, dropout with 50\% of probability is applied to the internal gates. Training was performed using Keras\footnote{\url{https://keras.io/}} with TensorFlow\footnote{\url{https://www.tensorflow.org/}} backend, categorical cross-entropy as loss function and Adam as optimizer.\\
Table~\ref{table:model} summarizes the structure of the NN and gives an overview of the parameter count for each layer. Here $N_w$, is the number of windows extracted from the input signal, and it depends on its length. Overall, the full network counts 194,596 parameters. If all the weights are represented as 8-bits fixed point numbers, the total space occupied in memory is slightly less than 200~KB, which is far below the size of the on-chip FLASH memory of the target platform. In summary, compared to the original network, we reduced the input window size from 512 to 256 samples, reduced the depth of last three layers, and replaced of the LSTM with a less complex GRU.
\begin{center}
\begin{table}
\caption{Architecture of the NN and parameter count\label{table:model}}
\hfill{}
\begin{tabular}{lcc}
\toprule
{Layer}&{Output shape}&{Parameter count}\\
\midrule
Input & ($N_w$, 256, 1) & -\\
Conv1 & ($N_w$, 128, 8) & 48   \\
Conv2 & ($N_w$, 64, 16) & 656  \\
Conv3 & ($N_w$, 32, 32) & 2592 \\
Conv4 & ($N_w$, 16, 64) &10,304\\
Conv5 & ($N_w$, 8, 64) & 20,544 \\
Conv6 & ($N_w$, 4, 128) & 41,088\\
Conv7 & ($N_w$, 2, 128) & 82,048\\
Global Average Pooling & ($N_w$, 128) & 0\\
GRU & (64) & 37,056\\
Dense+Softmax & (4) & 260\\
\bottomrule
\end{tabular}
\hfill{}
\end{table}
\end{center}
\subsection{Quantization}
\label{sect:quantization}
We quantized inputs, weights, and intermediate activations as 8-bits fixed point numbers in order to minimize the memory footprint and to maximize the speedup coming from SIMD instructions. Once the number of bits is fixed, we also require to determine the most appropriate quantization scheme. The naive approach is to allocate as many bits for the integer part as necessary to cover the whole range $[min,max]$, where min and max are respectively the minimum and maximum element in the set of weights. If, on the one hand, this prevents from cutting out too large or too small weights, on the other it might lead to allocate most of the bits for the integer part, thus losing resolution on the fractional part. If only few weights have values close to the border of the interval, it might be unnecessary to pay such a price~\cite{Lin2016}.

Our approach is to calculate mean and standard deviations of all weights, and then select the number of bits to allocate for the integer part in such a way that it could be possible to represent all values in the range $[\mu+3\sigma,\mu-3\sigma]$, where $\mu$ and $\sigma$ are respectively mean and standard deviation. Following this approach, we have opted for 2 integer bits and 5 fractional bits ($Q2.5$ notation). To verify that performance drops after quantization is acceptable, we first applied the transformations described in~\eqref{eq:quant}, then again divided by $2^5$. Thus, the resulting weights are fixed point numbers inside the representable range and with a granularity of $2^{-5}=0.03125$. To simulate the effect of quantization on intermediate activations, we inserted quantization layers after each pair convolution-pooling and at the output of the GRU. GRU's internal gates are quantized as 16-bits numbers in Q2.13 format. For that reason, we neglected the effect of the quantization on the aforementioned gates and internal states.

%%%%%%%%%%%%%%%%%%%%%%%%%%%%%%%%%%%%%%%%%%%%%%%%%%%%%%%%%%%%%%%%%%%%%%%%%%%%%%%%
\section{Results and Discussion}\label{sec:results}
In this section, we evaluate our approach in two steps. First, we assess the impact on the accuracy due to the modified architecture and quantization. Then, we evaluate the performance in terms of memory footprint, execution time, and overall operation count and throughput directly on the nRF52832. For all the experiments on the SoC, we built the firmware with the GNU Arm Embedded Toolchain~\footnote{\url{https://developer.arm.com/tools-and-software/open-source-software/developer-tools/gnu-toolchain/gnu-rm}} with level 3 optimization ($-O3$ flag on compilation command). In order to measure the execution time, we used the readout of the \texttt{CYCCNT} register.

\subsection{Accuracy}
After training for 250 epochs, the accuracy of the full precision network was 89.3\% on the training set and 86.1\% on the test set. The fixed point (FP) implementation achieved an accuracy of 85.7\%. Moreover, in Table~\ref{table:results}, we report the sensitivity (ratio of positives that are correctly detected), specificity (ratio of correctly detected negatives), and $F1$ score (harmonic mean between precision and recall) for each class and for the overall network.

\newcommand{\specialcell}[2][c]{\begin{tabular}[#1]{@{}c@{}}#2\end{tabular}}

\begin{center}
\begin{table}[h]
\caption{Accuracy metrics of the neural networks\label{table:results}}
\hfill{}
\begin{tabular}{@{}llccc@{}}
\toprule
Class & Metric & \specialcell{Training\\{set}} & \specialcell{{Test}\\{set}} & \specialcell{{Test set}\\{FP precision}} \\
\midrule
\multirow{3}{*}{Normal rhythm} & Sensitivity & 0.959 & 0.931 & 0.923 \\
& Specificity & 0.870 & 0.865 & 0.867 \\
& $F_1$ score & 0.936 & 0.920 & 0.916 \\
\addlinespace
\multirow{3}{*}{Atrial fibrillation} & Sensitivity & 0.864 & 0.841 & 0.848 \\
& Specificity & 0.982 & 0.969 & 0.965\\
& $F_1$ score & 0.841 & 0.776 & 0.765\\
\addlinespace
\multirow{3}{*}{Other rhythm} & Sensitivity & 0.795 & 0.741 & 0.745\\
& Specificity & 0.953 & 0.932 & 0.922\\
& $F_1$ score & 0.832 & 0.777 & 0.770\\
\addlinespace
\multirow{3}{*}{Noise} & Sensitivity & 0.627 & 0.706 & 0.588\\
& Specificity & 0.995 & 0.991 & 0.996\\
& $F_1$ score & 0.707 & 0.727 & 0.670\\
\addlinespace
\multirow{3}{*}{Overall} & Sensitivity & 0.811 & 0.805 & 0.776\\
&Specificity & 0.950 & 0.939 & 0.938\\
&F1 Score & 0.829 &  0.800 & 0.780 \\
&Accuracy & 0.893 & 0.861 & 0.854 \\ 
\bottomrule
\end{tabular}
\hfill{}
\end{table}
\end{center}
The last column of Table~\ref{table:results} summarizes the performance figures obtained with the modifications described in section \ref{sect:quantization} to simulate quantization. Sensitivity to noise is the most penalized, but except from that, performance metrics are not remarkably impacted by the fixed-point quantization, and in some case it even shows a slight improvement (\textit{e.g.}, Atrial Fibrillation sensitivity).

\begin{figure}
\centering
\includegraphics[width=0.9\linewidth]{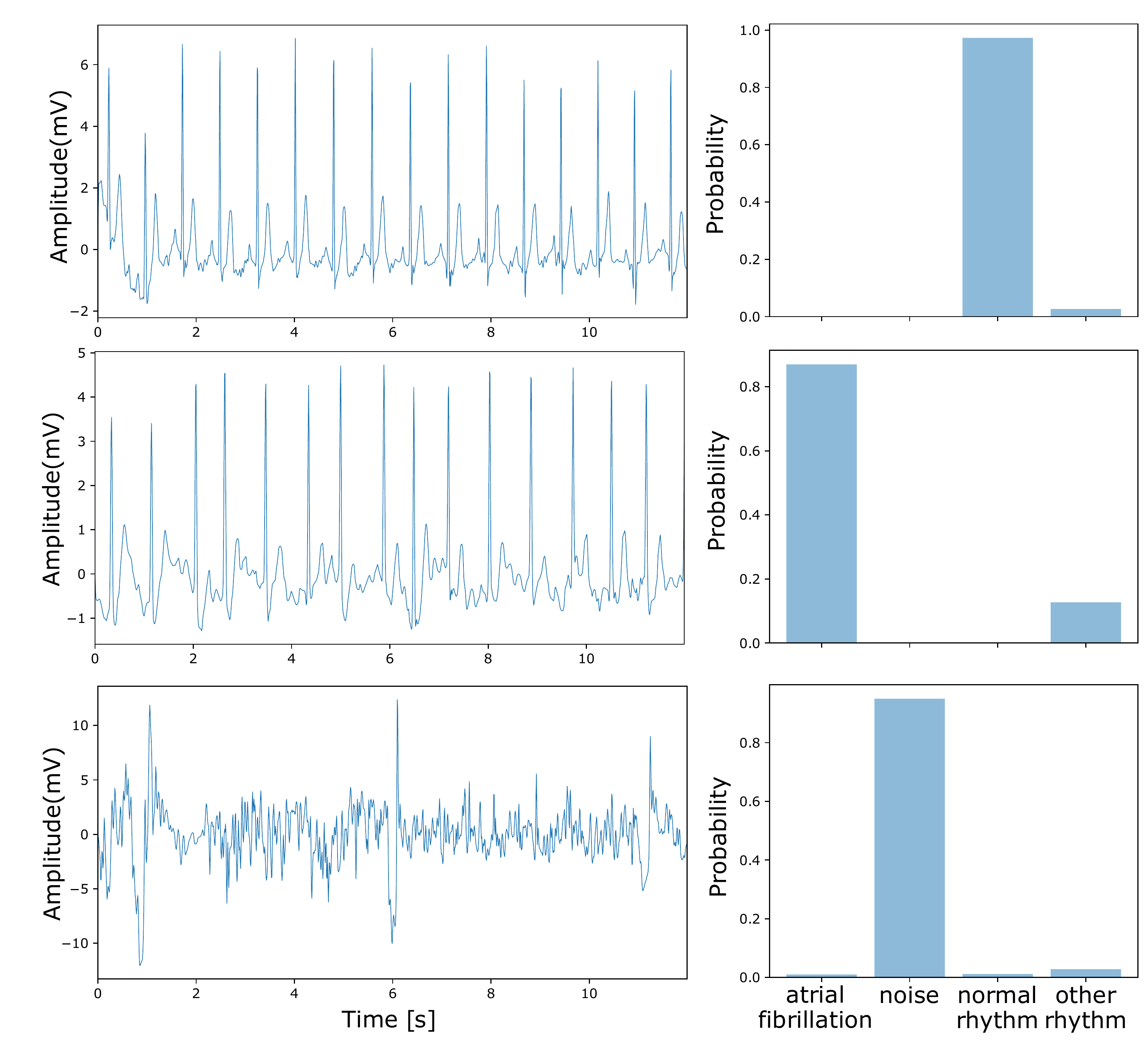}
\caption{ECG signals in the test set (left) and their corresponding class probabilities (right) estimated by the quantized NN. (top)~normal, (middle)~atrial fibrillation, (bottom)~noise.}
\end{figure}

\subsection{Memory Footprint}
The output binary is around 210~KB, which include weights and biases, the routines of CMSIS-DSP and CMSIS-NN necessary to run the network, and few other lines of code for setup and configuration of the board. By looking at the \texttt{.map} file generated by the toolchain, we found that the memory allocated for hard-coded data (namely weights, biases and lookup tables for trigonometric functions) is 195.6~KB. Data allocated in RAM consists mainly in GRU gates and hidden states (0.8~KB), buffers for convolutions (4~KB), and convolutions activations (two arrays of 2~KB).

\subsection{Timing}
In order to estimate the execution time, we fed the network 4 windows of data (640 ECG samples), which corresponds to 4 inferences of the NN. Then, we calculated the difference between the value stored into the \texttt{CYCCNT} register before and after the execution of the network. By dividing the obtained number by 4, we obtain the estimated average execution time. With the above configuration, we obtained an interval of 379.2~ms, which translates into an average processing time per window of 94.8~ms. The largest part of this interval, around 91~ms, is spent during the convolutional part of the network, 3.8~ms are spent during the execution of the GRU, and 28~$\mu$s are spent in the fully connected layer.

\subsection{Operation Count and Throughput}
We based our estimations on the following assumptions in order to obtain an estimation of the number of operations that the network performs to process one window:
\begin{itemize}
\item For 1D convolutional layers, we assume $2*K*C*N*L+L*N$ operations, where K is the kernel size, C the input channels, N the output channels, and L the length of the output of the layer. The second addend is the contribution from the biases.
\item Averaging pooling layers with kernel dimension of 2 and stride 2 amount $L*C/2$ operations.
\item Fully connected layers amount to $2*M*N+N$ operations, where $M$ and $N$ are the dimension of the input and the output tensor respectively.
\item We neglect the contributions of activation functions since in CMSIS these transformations are implemented as lookup tables or bitwise operations.
\end{itemize}
The total number of operations of the network under these assumptions is summarized in Table~\ref{table:opscount}. Finally, we calculate the throughput of our implementation as
$$\mathrm{Throughput}=\frac{\mathrm{OpsCount}}{\mathrm{ExecutionTime}}=33.98~\mathrm{MOps/s},$$
where the operation count and the execution time are constrained to the execution of a single window. Moreover, since the clock frequency of our SoC is 64~MHz, this translates into an average of $0.531~\mathrm{Ops/Cycle}$.

\begin{center}
\begin{table}[h]
\caption{Estimated operation counts of the network\label{table:opscount}}
\hfill{}
\begin{tabular}{lcc}
\toprule
Layer & Parameter count & Operation count\\
\midrule
Convolutional block & 157,280 & 3.147~MOps\\
GRU & 37,056 & 73,728~KOps\\
Dense layer & 260 & 516~Ops\\
Total & 194,596 & 3.221~MOps\\
\bottomrule
\end{tabular}
\hfill{}
\end{table}
\end{center}

\subsection{Power and Efficiency}
We calculated the current consumption of the system by measuring the voltage drop on the $33~\Omega$ resistor in series to the supply line. The board is powered with 5V, the DC/DC converter is enabled, and the processor executes the network continuously in a loop. We measured a voltage drop of 136.25~mV, which translates into an input current of 4.13~mA and a power of 20.65~mW. We finally, calculated power efficiency as
$$\frac{\mathrm{Throughput}}{\mathrm{Power}}=\frac{33.98~\mathrm{MOps/s}}{20.65~\mathrm{mW}}=1.64~\mathrm{GOps/s/W}.$$
In a final implementation, SoC is idle for most of the time, thus the RMS must be calculated according to the duty cycle. We replicated the same measurements using TensorFlow Lite for Microcontrollers as inference engine. With 8-bits quantization, for the only convolutional part (RNNs are currently not supported) the efficiency is just $\frac{\mathrm{Throughput}}{\mathrm{Power}}=\frac{3.0~\mathrm{MOps/s}}{24.14~\mathrm{mW}}=0.124~\mathrm{GOps/s/W}$. 

%%%%%%%%%%%%%%%%%%%%%%%%%%%%%%%%%%%%%%%%%%%%%%%%%%%%%%%%%%%%%%%%%%%%%%%%%%%%%%%%
\section{Conclusion}
We presented a NN for arrhythmia detection that, in terms of size and computational complexity, is suitable for deployment to a resource-constrained microcontroller. To achieve so, we expressed weights and activations as 8-bits integers in $Q$ format. We then implemented such network on our target platform using CMSIS-NN and benchmarked it (memory footprint, execution time, and throughput). In future works, we will perform a detailed comparison between different inference libraries, including TensorFlow Lite for Microcontrollers and different hardware platforms and accelerators like the GAP8 from GreenWaves Technologies~\footnote{\url{https://greenwaves-technologies.com/ai_processor_gap8/}}. The best performing solution will eventually be integrated in a wearable device that acquires and processes ECG signals in real-time.

%%%%%%%%%%%%%%%%%%%%%%%%%%%%%%%%%%%%%%%%%%%%%%%%%%%%%%%%%%%%%%%%%%%%%%%%%%%%%%%%
\bibliography{references}

\begin{thebibliography}{10}
\providecommand{\url}[1]{#1}
\csname url@samestyle\endcsname
\providecommand{\newblock}{\relax}
\providecommand{\bibinfo}[2]{#2}
\providecommand{\BIBentrySTDinterwordspacing}{\spaceskip=0pt\relax}
\providecommand{\BIBentryALTinterwordstretchfactor}{4}
\providecommand{\BIBentryALTinterwordspacing}{\spaceskip=\fontdimen2\font plus
\BIBentryALTinterwordstretchfactor\fontdimen3\font minus
  \fontdimen4\font\relax}
\providecommand{\BIBforeignlanguage}[2]{{%
\expandafter\ifx\csname l@#1\endcsname\relax
\typeout{** WARNING: IEEEtran.bst: No hyphenation pattern has been}%
\typeout{** loaded for the language `#1'. Using the pattern for}%
\typeout{** the default language instead.}%
\else
\language=\csname l@#1\endcsname
\fi
#2}}
\providecommand{\BIBdecl}{\relax}
\BIBdecl

\bibitem{Miotto2017}
R.~Miotto, F.~Wang, S.~Wang, X.~Jiang, and J.~T. Dudley, ``Deep learning for
  healthcare: review, opportunities and challenges,'' \emph{Briefings in
  bioinformatics}, vol.~19, no.~6, pp. 1236--1246, 2017.

\bibitem{Dunn2018}
J.~Dunn, R.~Runge, and M.~Snyder, ``Wearables and the medical revolution,''
  \emph{Personalized medicine}, vol.~15, no.~5, pp. 429--448, 2018.

\bibitem{Nelson2016}
E.~C. Nelson, T.~Verhagen, and M.~L. Noordzij, ``Health empowerment through
  activity trackers: an empirical smart wristband study,'' \emph{Computers in
  human behavior}, vol.~62, pp. 364--374, 2016.

\bibitem{Schneegass2017}
S.~Schneegass and O.~Amft, \emph{Smart textiles}.\hskip 1em plus 0.5em minus
  0.4em\relax Springer, 2017.

\bibitem{Mombers2016}
C.~Mombers, K.~Legako, and A.~Gilchrist, ``Identifying medical wearables and
  sensor technologies that deliver data on clinical endpoints,'' \emph{British
  Journal of Clinical Pharmacological}, vol.~81, no.~2, pp. 196--198, Feb.
  2016.

\bibitem{Rajpurkar2017}
P.~Rajpurkar, A.~Y. Hannun, M.~Haghpanahi, C.~Bourn, and A.~Y. Ng,
  ``Cardiologist-level arrhythmia detection with convolutional neural
  networks,'' \emph{arXiv Preprint}, 2017.

\bibitem{Chen2017}
Z.~Chen, W.~Hu, J.~Wang, S.~Zhao, B.~Amos, G.~Wu, K.~Ha, K.~Elgazzar,
  P.~Pillai, R.~Klatzky, D.~Siewiorek, and M.~Satyanarayanan, ``An empirical
  dtudy of latency in an emerging class of edge computing applications for
  wearable cognitive assistance,'' in \emph{Proceedings of the 2nd {ACM/IEEE}
  Symposium on Edge Computing}, 2017, pp. 1--14.

\bibitem{Rashid2017}
H.~Rashid, I.~U. Ahmed, R.~Das, and S.~M.~T. Reza, ``Emergency wireless health
  monitoring system using wearable technology for refugee camp and disaster
  affected people,'' in \emph{Proceedings of the International Conference on
  Computer, Communication, Chemical, Materials and Electronic Engineering
  ({IC4ME2'17})}, 2017, pp. 144--147.

\bibitem{Horvitz2015}
E.~Horvitz and D.~Mulligan, ``Data, privacy, and the greater good,''
  \emph{Science}, vol. 349, no. 6245, pp. 253--255, 2015.

\bibitem{Lee2013}
K.~H. Lee and N.~Verma, ``A low-power processor with configurable embedded
  machine-learning accelerators for high-order and adaptive analysis of
  medical-sensor signals,'' \emph{{IEEE} Journal of Solid-State Circuits},
  vol.~48, no.~7, pp. 1625--1637, Jul. 2013.

\bibitem{Li2018}
H.~Li, K.~Ota, and M.~Dong, ``Learning {IoT} in edge: deep learning for the
  {Internet of Things} with edge computing,'' \emph{IEEE Network}, vol.~32,
  no.~1, pp. 96--101, 2018.

\bibitem{Linares-Barranco2019}
A.~Linares-Barranco, A.~Rios-Navarro, R.~Tapiador-Morales, and T.~Delbruck,
  ``Dynamic vision sensor integration on {FPGA}-based {CNN} accelerators for
  high-speed visual classification,'' \emph{arXiv Preprint}, 2019.

\bibitem{Gao2019}
C.~Gao, S.~Braun, I.~Kiselev, J.~Anumula, T.~Delbruck, and S.~Liu, ``Real-time
  speech recognition for {IoT} purpose using a delta recurrent neural network
  accelerator,'' in \emph{Proceedings of the 2019 IEEE International Symposium
  on Circuits and Systems ({ISCAS'19})}, May 2019, pp. 1--5.

\bibitem{Ignatov2018}
A.~Ignatov, R.~Timofte, W.~Chou, K.~Wang, M.~Wu, T.~Hartley, and L.~Van~Gool,
  ``{AI} benchmark: running deep neural networks on {Android} smartphones,'' in
  \emph{Workshops The European Conference on Computer Vision ({ECCV'18})},
  September 2018.

\bibitem{Bennett2012}
D.~H. Bennett, \emph{Bennett's cardiac arrhythmias: practical notes on
  interpretation and treatment}.\hskip 1em plus 0.5em minus 0.4em\relax John
  Wiley \& Sons, 2012.

\bibitem{VanZaen2019}
J.~Van~Zaen, O.~Chételat, M.~Lemay, E.~Muntané~Calvo, and R.~Delgado-Gonzalo,
  ``Classification of cardiac arrhythmias from single lead {ECG} with a
  convolutional recurrent neural network,'' in \emph{Proceedings of the 12th
  International Joint Conference on Biomedical Engineering Systems and
  Technologies ({BIOSTEC'19})}, vol.~4, Jan. 2019, pp. 33--41.

\bibitem{Clifford2017}
G.~D. Clifford, C.~Liu, B.~Moody, H.~L. Li-wei, I.~Silva, Q.~Li, A.~E. Johnson,
  and R.~G. Mark, ``{AF} classification from a short single lead {ECG}
  recording: the {PhysioNet/Computing in Cardiology Challenge 2017},'' in
  \emph{Proceedings of the Computing in Cardiology ({CinC'17})}.\hskip 1em plus
  0.5em minus 0.4em\relax IEEE, 2017, pp. 1--4.

\bibitem{Chetelat2015}
O.~Chételat, D.~Ferrario, M.~Proença, J.-A. Porchet, A.~Falhi,
  O.~Grossenbacher, R.~Delgado-Gonzalo, N.~Della~Ricca, and C.~Sartori,
  ``Clinical validation of {LTMS-S}: a wearable system for vital signs
  monitoring,'' in \emph{Proceedings of the 37th Annual International
  Conference of the {IEEE} Engineering in Medicine and Biology Society
  ({EMBC'15})}, 2015, pp. 3125--3128.

\bibitem{Lai2018}
L.~Lai, N.~Suda, and V.~Chandra, ``{CMSIS-NN:} efficient neural network kernels
  for {Arm} {Cortex-M} {CPUs},'' \emph{arXiv Preprint}, 2018.

\bibitem{Lin2016}
D.~Lin, S.~Talathi, and S.~Annapureddy, ``Fixed point quantization of deep
  convolutional networks,'' in \emph{International Conference on Machine
  Learning}, 2016, pp. 2849--2858.

\end{thebibliography}
\theendnotes

%%%%%%%%%%%%%%%%%%%%%%%%%%%%%%%%%%%%%%%%%%%%%%%%%%%%%%%%%%%%%%%%%%%%%%%%%%%%%%%%
\end{document}